\documentclass[aps,pra,twocolumn,showpacs,preprintnumbers,amsmath,amssymb,footinbib]{revtex4}
\usepackage{mathtools}

\DeclarePairedDelimiter{\floor}{\lfloor}{\rfloor}
\usepackage{braket}
\usepackage{bbold}
\usepackage{graphicx,epsfig}
\usepackage{bm}
\usepackage{dcolumn}
\usepackage{xcolor}
\usepackage[breaklinks=true,colorlinks,citecolor=blue,linkcolor=blue,urlcolor=blue]{hyperref}

\def\sc#1#2#3{Science {\bf #1}, #2 (#3)}
\def\prv#1#2#3{Phys. Rev. {\bf #1}, #2 (#3)}
\def\rmp#1#2#3{Rev. Mod. Phys. {\bf #1}, #2 (#3)}
\def\prl#1#2#3{Phys. Rev. Lett. {\bf #1}, #2 (#3)}
\def\pra#1#2#3{Phys. Rev. A {\bf #1}, #2 (#3)}

\def\epjd#1#2#3{Eur. Phys. J. D {\bf #1}, #2 (#3)}
\def\njp#1#2#3{New J. Phys. {\bf #1}, #2 (#3)}
\def\jstat#1#2#3{J. Stat. Mech. {\bf #1}, #2 (#3)}
\def\jpamt#1#2#3{J. Phys. A: Math. Theor. {\bf #1}, #2 (#3)}

\def\jpb#1#2#3{J. Phys. B: At. Mol. Opt. Phys. {\bf #1}, #2 (#3)}

\def\pla#1#2#3{Phys. Lett. A {\bf #1}, #2 (#3)}

\def\ajp#1#2#3{Am. J. Phys. {\bf #1}, #2 (#3)}
\def\ejp#1#2#3{Eur. J. Phys. {\bf #1}, #2 (#3)}

\def\noi{\noindent}
\def\bc{\begin{center}}
\def\ec{\end{center}}
\newcommand{\bea}{\begin{equation}}
\newcommand{\eea}{\end{equation}\noi}
\newcommand{\ber}{\begin{eqnarray}}
\newcommand{\eer}{\end{eqnarray}\noi}

\begin{document}
\title{Finite-size effects on the cluster expansions for quantum gases in restricted geometries}

\author{Soumi Dey\footnote{Present Address: Department of Physics, IIT-Guwahati, Guwahati-781039, India}}
\author{Prathyush Manchala}
\author{Srijit Basu\footnote{Present Address: Same as above}}
\author{Debshikha Banerjee$^{2}$\footnote{Present Address: 5/28 Old CIT Building, Beleghata, Kolkata-700010, India}}
\author{Shyamal Biswas}\email{sbsp [at] uohyd.ac.in}

\affiliation{School of Physics, University of Hyderabad, C.R. Rao Road, Gachibowli, Hyderabad-500046, India
}

\date{May 8, 2020}

\begin{abstract}
We have analytically obtained 1-particle density matrices for ideal Bose and Fermi gases in both the 3-D box geometries and the harmonically trapped geometries for the entire range of temperature. We have obtained quantum cluster expansions of the grand free energies in closed forms for the same systems in the restricted geometries. We have proposed a theorem (with a proof) about the generic form of the quantum cluster integral. We also have considered short ranged interactions in our analyses for the quasi 1-D cases of Bose and Fermi gases in the box geometries. Our theoretical results are exact, and are directly useful for understanding finite-size effects on quantum cluster expansion of Bose and Fermi gases in the restricted geometries. Our results would be relevant in the context of experimental study of spatial correlations in ultra-cold systems of dilute Bose and Fermi gases of alkali atoms (i) in 3-D magneto-optical box traps with quasi-uniform potential around the center \cite{Gaunt}, and (ii) in 3-D harmonic traps \cite{Ensher,Jin}.
\end{abstract}


\maketitle
 

\section{Introduction}
Density matrix is of very high interest in physics \cite{Neumann,Landau2,Dirac,Tolman,Feynman}. It maps equilibrium statistical mechanics to quantum dynamics and vice versa with the application of Wick rotation ($1/k_BT\leftrightarrows it/\hbar$) which maps inverse temperature ($1/T$) to imaginary time ($it$) \cite{Wick,Feynman}. Density matrix elements, which physically represents spatial correlations in a thermodynamic or mechanical system, are nothing but the propagators in the position representation \cite{Feynman}. Density matrix elements are thus useful to get path integrals in statistical field theory and quantum field theory \cite{Feynman}. 

Density matrix is introduced in several branches of physics such as statistical mechanics, many-body physics, quantum mechanics, quantum field theory, quantum optics, \textit{etc} \cite{Feynman,Fox}. Density matrix elements (and 1-particle density matrix elements for many-body systems \cite{Penrose,Pitaevskii}) are often calculated in the position representation for the class for systems having no boundaries at all, e.g. a free particle, a harmonic oscillator, a free Bose gas at a temperature $T$, a free Fermi gas at a temperature $T$, \textit{etc} \cite{Feynman,Barragan-Gila}. There have been many discussions in the frontiers level in connection with the 1-particle density matrices, cluster expansions, and spatial correlations in quantum gases \cite{Onofri,Batchelor,Gangardt,Caux,Kira}. 

However, hardly any discussions are found on the finite-size effects on the density matrices and quantum cluster expansions. Practically all the thermodynamic systems which come to equilibrium with the respective heat  (and in some occasions particle) reservoirs are bounded. Hence we are interested in calculating 1-particle density matrix \cite{Pitaevskii} for ideal Bose and Fermi gases in 3-D box geometries at a temperature $T$. We are also interested to explore the same for interacting Bose and Fermi gases confined at least in quasi 1-D boxes.  We also want to obtain quantum cluster expansions \cite{Khan,Pathria} of the grand free energies for these systems in connection with the density matrices. Thermodynamic properties of the systems and the finite-size effects on them can be easily obtained from the quantum cluster expansions of the grand free energies.

Study of ultracold quantum gases has been a topic of high experimental and theoretical interest \cite{Dalfovo,Bloch,Giorgini} after the observation of Bose-Einstein condensation of alkali atoms in 3-D magneto-optical harmonic traps in 1995 \cite{Anderson,Bradley,Davis}. Bose-Einstein condensation of a dilute Bose gas of $^{87}$Rb atoms has also been observed recently in a 3-D magneto-optical box trap with quasi-uniform potential in it \cite{Gaunt}. Finite-size effect has also been observed on harmonically trapped Bose gas in the form of (i) shift of the condensation point \cite{Ensher}, (ii) collapse of the condensate (for attractive interactions) \cite{Roberts}, (iii) Casimir-Polder effect on the condensate near a substrate \cite{Harber}, (iv) remnant of zero-point energy effects \cite{Castilho}, \textit{etc}. Thus, studying 1-particle density matrix for ideal Bose and Fermi gases in 3-D box geometries and that of quantum cluster expansions of these systems would be relevant in the current context.  Surprisingly, any discussions on the quantum cluster expansion are also not found even for the harmonically trapped Bose or Fermi gas which drew a lot of experimental and theoretical interests since 1995 \cite{Anderson,Jin,Dalfovo,Bloch,Giorgini,Fetter}. Hence, studying the quantum cluster expansions of the Bose and Fermi gases in harmonically trapped geometries would be relevant in the current context.

Finite-size effects have already been theoretically studied on quantum gases \cite{Pathria2,Subrahmanyam,Bhattacharyya,Brankov,Biswas2007}, specially on (i) the respective shifts of the Bose-Einstein condensation point and the Fermi temperature of a harmonically trapped ultracold Bose gas \cite{Grossmann,Ketterle} and Fermi gas \cite{Biswas2012}, (ii) Casimir-Polder effect on a harmonically trapped Bose-Einstein condensate near a substrate \cite{Antezza}, (iii) Casimir effect on a Bose-Einstein condensate in slabs \cite{SB}, (iv) collapse of attractively interacting Bose-Einstein condensate \cite{Baym,Biswas2009}, \textit{etc}. However, our work on the finite-size effects on the cluster expansion for quantum gases in restricted geometries is absolutely novel. Our method of obtaining the quantum cluster expansion, with the realization of the quantum cluster integral as the partition function of a composite particle and having a phase, would be a rigorous approach for studying finite-size effects in statistical mechanics.

Calculations in this article begin with the introduction of the statistical mechanical density matrix and the equation for its matrix-element in position-space representation for a single particle in equilibrium with a heat bath at an absolute temperature $T$. Then we obtain the density matrix-element for a particle in 1-D box. We generalize the density matrix-element for a single particle in a 3-D box. Then we generalize the 3-D case of the single particle to the ideal gases of many-particles (i.e. for indistinguishable bosons and fermions) with the 1-particle density matrix within grandcanonical ensemble. We obtain quantum cluster expansions for grand free energies for these systems. Then we consider the cases of short ranged interactions in our analyses specially for the quasi 1-D cases of Bose and Fermi gases in the box geometries. We further study the quantum cluster expansions for the Bose and Fermi gases in harmonically trapped geometries. Finally, we conclude along with a discussion on the theorem (with a proof in the appendix) about the generic form of the quantum cluster integral. 

\section{Statistical Mechanical Density Matrix}
Let us consider a thermodynamic system in equilibrium with a heat bath at an absolute temperature $T$. Statistical mechanical density matrix is defined for the system as \cite{Feynman}
\begin{eqnarray}\label{eqn:1}
\hat{\rho}=\text{e}^{-\beta\hat{H}}
\end{eqnarray}
where $\hat{H}$ is the Hamiltonian of the system, $\beta=1/k_BT$ and $k_B$ is the Boltzmann constant. Though a normalization factor $1/\text{Tr}.\hat{\rho}$ is needed to normalize the density matrix, we are not considering it in the definition. Inverse of the normalization factor is called the partition function ($Z=\text{Tr}.\hat{\rho}$ \cite{Feynman}), which can be separately considered whenever needed to extract thermodynamic properties of the system. Let $\{|\psi_j\rangle\}; j=1,2,3,....$ be the orthonormalized ($\langle\psi_i|\psi_j\rangle=\delta_{i,j}$ $\forall~i,j$) and complete set of energy eigenstates ($\sum_{j=1}^\infty|\psi_j\rangle\langle\psi_j|=\mathbb{1}$) of the system with the respective eigenvalues $\{E_j\}$, then density matrix in Eqn.(\ref{eqn:1}) can be recast as \cite{Feynman}  
\begin{eqnarray}\label{eqn:2}
\hat{\rho}=\sum_{j=1}^\infty\text{e}^{-\beta E_j}|\psi_j\rangle\langle\psi_j|
\end{eqnarray}

\begin{figure}
\includegraphics[width=0.98 \linewidth]{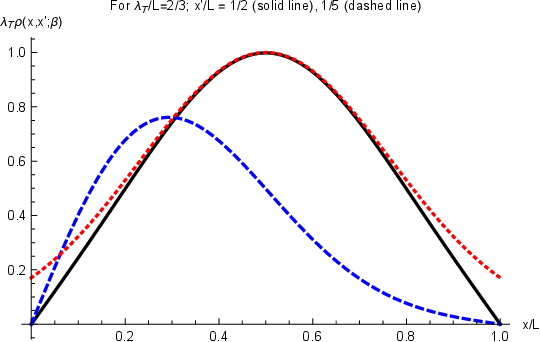}
\caption{Profile of the density matrix elements for the particle in the 1-D box of length $L$. The solid and dashed lines follow Eqn. (\ref{eqn:8}) for $x'/L=1/2$ and $1/5$ respectively. The dotted line represents the boundary-free case, and follows Eqn. (\ref{eqn:5}) for $x'/L=1/2$. Thermal de Broglie wavelength of the system, $\lambda_T=\sqrt{\frac{2\pi\hbar^2\beta}{m}}$, is set as $\lambda_T=2L/3$ for all the plots.
\label{fig1}}
\end{figure}

\subsection{Density matrix for a single particle in a 1-D box}
If the points $x'$ and $x$ are any two arbitrary points inside a 1-D box of length $L$ which confines a point-particle such that $0<x'<L$ and $0<x<L$, then the density matrix element of the system can be defined in position representation as  
\begin{eqnarray}\label{eqn:3}
\rho(x,x';\beta)=\langle x|\hat{\rho}|x'\rangle&=&\langle x|\text{e}^{-\beta\hat{H}}|x'\rangle\nonumber\\&=&\sum_{j=1}^\infty\text{e}^{-\beta E_j}\psi_j(x)\psi_j^*(x').
\end{eqnarray}
Since energy eigenstate $\psi_j(x)$ takes the form $\psi_j(x)=\sqrt{\frac{2}{L}}\sin(j\pi x/L)$ for the particle in the 1-D box, we can recast Eqn.(\ref{eqn:3}) as
\begin{eqnarray}\label{eqn:7}
\rho(x,x';\beta)&=&\sum_{j=1}^\infty\frac{2}{L}\text{e}^{-\beta\frac{j^2\pi^2\hbar^2}{2mL^2}}\sin\big(\frac{j\pi x'}{L}\big)\sin\big(\frac{j\pi x}{L}\big)
\end{eqnarray}
The above summation over $j$ can be replaced by the integral over $j$ for $L\rightarrow\infty$ to get the free particle result for the density matrix element, counter to Eqn.(\ref{eqn:7}) as for $x$ and $x'$ at around $L/2$, as
\begin{eqnarray}\label{eqn:5}
\rho_f(x,x';\beta)=\sqrt{\frac{m}{2\pi\hbar^2\beta}}\text{e}^{-\frac{m(x'-x)^2}{2\hbar^2\beta}}
\end{eqnarray}
This result, for the free particle in the thermal equilibrium, is well known in the literature \cite{Feynman}. 

However, exact evaluation of the summation in Eqn.(\ref{eqn:7}) is not obvious, and it has not been exactly evaluated for a particle in the box as far as we know. Replacing $\big[2\sin\big(\frac{j\pi x'}{L}\big)\sin\big(\frac{j\pi x}{L}\big)\big]$ by $\big[\cos\big(\frac{j\pi[x'-x]}{L}\big)-\cos\big(\frac{j\pi[x'+x]}{L}\big)\big]$ and performing the summation over $j$ in Eqn.(\ref{eqn:7}), we get an exact result for the density matrix element, in compact form, as
\begin{eqnarray}\label{eqn:8}
\rho(x,x';\beta)&=&\frac{1}{2L}\bigg[\vartheta_3\bigg(\frac{\pi[x'-x]}{2L},\text{e}^{-\beta\frac{\pi^2\hbar^2}{2mL^2}}\bigg)\nonumber\\&&-\vartheta_3\bigg(\frac{\pi[x'+x]}{2L},\text{e}^{-\beta\frac{\pi^2\hbar^2}{2mL^2}}\bigg)\bigg]
\end{eqnarray}
where $\vartheta_3$ represents the Jacobi (elliptic) theta function of the 3rd kind in the usual notation\footnote{The Jacobi (elliptic) theta function, $\vartheta_3(u,q)$, is defined as $\vartheta_3(u,q)=1+2\sum_{j=1}^\infty q^{j^2}\cos(2ju)$ for $u\in\mathbb{C}$ and $|q|<1$.}. This is a new result, as the summation in the preceding equation, though looks simple, was not evaluated before us in the context of the density matrix for a particle in a 1-D box. The second term in the right hand side of Eqn.(\ref{eqn:8}) breaks the translational symmetry in the density matrix-element, as because, the density matrix-element no longer depends on separation of the points $x$ and $x'$ for finite $L$. The translation symmetry, however, is preserved for the free particle as clear in the Eqn.(\ref{eqn:5}). We plot the density matrix-element (Eqn.(\ref{eqn:8})) and also compare it with the one with the free-boundary (Eqn.(\ref{eqn:5})) in FIG. \ref{fig1}. Eqn.(\ref{eqn:8}) is significantly different from Eqn.(\ref{eqn:5}) in the low temperature regime when thermal de Broglie wavelength, $\lambda_T=\sqrt{\frac{2\pi\hbar^2}{mk_BT}}$, becomes comparable to or bigger than the system size ($L$). It is also clear from the dashed line of the figure, that, the density matrix-element is not being maximized at $x'=x$ unless $x'=L/2$\footnote{Maximization of the density matrix-element is possible at $x'=x=L/2$ as there is a reflection symmetry about this point.}, as the translation symmetry is broken for the finite-size ($L$) of the system.

\subsection{Density matrix for a single particle in a 3-D box}
Let us now consider the particle to be in a 3-D box of lengths $L_1, L_2, L_3$ along the $x,y,z$ axes respectively. Position of the particle is now given by $\vec{r}=x\hat{i}+y\hat{j}+z\hat{k}$ (or $\vec{r}'=x'\hat{i}+y'\hat{j}+z'\hat{k}$) such that $0<x<L_1$, $0<y<L_2$, $0<z<L_3$. Since the system is linear, as there are no inter-particle interactions, motions of the particle along $x$, $y$ and $z$ axes would be independent of each other. Thus, we can generalize the result for the density matrix-element in Eqn.(\ref{eqn:8}) for the 3-D as product of the density matrix elements for the individual axes:
\begin{eqnarray}\label{eqn:9}
\rho(\vec{r},\vec{r}';\beta)&=&\Pi_{i=1}^3\frac{1}{2L_i}\bigg[\vartheta_3\bigg(\frac{\pi[x_i'-x_i]}{2L_i},\text{e}^{-\beta\frac{\pi^2\hbar^2}{2mL_i^2}}\bigg)\nonumber\\&&-\vartheta_3\bigg(\frac{\pi[x_i'+x_i]}{2L_i},\text{e}^{-\beta\frac{\pi^2\hbar^2}{2mL_i^2}}\bigg)\bigg]
\end{eqnarray}
where $x_1=x$, $x_2=y$, $x_3=z$, and so as for the primed coordinates. Eqn.(\ref{eqn:9}) is our result for statistical mechanical density matrix element ($\rho(\vec{r},\vec{r}';\beta)=\langle\vec{r}|\hat{\rho}|\vec{r}'\rangle$) for a single particle in a 3-D box in equilibrium with a heat bath at a temperature $T$. 

The forms of the Eqns.(\ref{eqn:8}) and (\ref{eqn:9}) reveal the insight of breaking of the translational symmetry of the density-matrix for a particle in the respective box geometries. However, translational symmetry breaking would not just be the property of the box geometry, rather would be the property of the finiteness of the geometry where the system is confined to. We can investigate the same for a many-body systems confined to restricted geometries not only for the box-geometries but also for the harmonically trapped geometries. In the following section, we extend the statistical mechanical density matrix for many-body systems within grandcanonical ensemble.

\section{One-particle density matrix for an ideal quantum gas in a 3-D box}
One-particle density matrix for an ideal quantum gas ( i.e. Bose or Fermi gas of indistinguishable particles) in equilibrium with a heat (and particle) bath at a temperature $T$ (and chemical potential $\mu$) is defined (by generalizing Eqn.(\ref{eqn:2})) as
\cite{Penrose,Pitaevskii}
\begin{eqnarray}\label{eqn:10}
\hat{\rho}^{1}=\sum_{j=1}^\infty\frac{1}{\text{e}^{\beta(E_j-\mu)}\pm1}|\psi_j\rangle\langle\psi_j|
\end{eqnarray}
where $\{|\psi_j\rangle\}$ are the orthonormalized and complete set of eigenstates of eigenvalues $\{|E_j\rangle\}$  for the single-particle Hamiltonian of the many-body system and the pre-factor, $\frac{1}{\text{e}^{\beta(E_j-\mu)}}$, represents the average occupation number of particles ($\bar{n}_j$) to the single-particle state $|\psi_j\rangle$.  Thus we generalize Eqn.(\ref{eqn:9})to the 1-particle density matrix for the many-body system in the 3-D box as
\begin{eqnarray}\label{eqn:11}
\rho^{1}(\vec{r},\vec{r}';\beta)&=&\sum_{j_1=1}^\infty\sum_{j_2=1}^\infty\sum_{j_3=1}^\infty\frac{1}{\text{e}^{\beta([E_{j_1}+E_{j_2}+E_{j_3}]-\mu)}\mp1}\nonumber\\&&\times\Pi_{i=1}^3\frac{2}{L_i}\sin\bigg(\frac{j_i\pi x_i}{L_i}\bigg)\sin\bigg(\frac{j_i\pi x_i'}{L_i}\bigg)~~~~~
\end{eqnarray}
where $E_{j_i}=\frac{\pi^2\hbar^2{j_i^2}}{2mL_i^2}$ for $i=1,2,3$, $E_{j_1, j_2, j_3}=E_{j_1}+E_{j_2}+E_{j_3}$ is the energy eigenvalue in the single-particle sate $\ket{\psi_{j_1, j_2, j_3}}$, upper sign represents Bose gas and lower sign represents Fermi gas.

While single-particle ground state energy of the system is given by $E_{1,1,1}=\frac{\pi^2\hbar^2}{2m}\big[\frac{1}{L_1^2}+\frac{1}{L_2^2}+\frac{1}{L_3^2}\big]$, fugacity of the system is given by $\bar{z}=\text{e}^{\mu/k_BT}$ which ranges as $0\le\bar{z}<\text{e}^{E_{1,1,1}/k_BT}$ for the ideal Bose gas and $0\le\bar{z}<\infty$ for the ideal Fermi gas. One-particle density matrix-element (in position representation) catches long-ranged order in the many-body system at least for 3-D free Bose gas \cite{Penrose,Pitaevskii}. The finite-size effect kills the long-ranged order in the many-body system as the density matrix element has to vanish at the boundaries as clear even in FIG. \ref{fig1}.  Expansion of the r.h.s. of Eqn.(\ref{eqn:11}) around $\bar{z}=0$ leads to 
\begin{eqnarray}\label{eqn:12}
\rho^{1}(\vec{r},\vec{r}';\beta)=\sum_{l=1}^\infty(\pm1)^{l-1}\bar{z}^{l}\rho(\vec{r},\vec{r}';l\beta).
\end{eqnarray}
The fugacity can be determined by fixing total average number of particles which can be directly evaluated by integrating the  diagonal elements of the 1-particle density matrix ($\bar{N}=\int\rho^{1}(\vec{r},\vec{r};\beta)\text{d}^3\vec{r}$), as
\begin{eqnarray}\label{eqn:12b}
\bar{N}&=&\sum_{l=0}^\infty(\pm1)^{l-1}\bar{z}^l\Pi_{i=1}^3\bigg[\frac{1}{2}\vartheta_3\bigg(0,\text{e}^{-\frac{l\beta\pi^2\hbar^2}{2mL_i^2}}\bigg)\nonumber\\&&-\frac{1}{2}\vartheta_3\bigg(\frac{\pi}{2},\text{e}^{-\frac{l\beta\pi^2\hbar^2}{2mL_i^2}}\bigg)-\sum_{n=1}^\infty\frac{\pi^{2n}}{2^{2n+1}(2n+1)!}\nonumber\\&&\times\vartheta_3^{'(2n-1)}\bigg(\frac{\pi}{2},\text{e}^{-\frac{l\beta\pi^2\hbar^2}{2mL_i^2}}\bigg)\bigg]
\end{eqnarray}
where $\big[\frac{\pi}{2}\big]^n\vartheta_3^{'(n-1)}\bigg(\frac{\pi}{2},\text{e}^{-\frac{l\beta\pi^2\hbar^2}{2mL_i^2}}\bigg)$ is the nth order derivative of the Jacobi (elliptic) theta function $\vartheta_3\bigg(\frac{\pi}{2}x,\text{e}^{-\frac{l\beta\pi^2\hbar^2}{2mL_i^2}}\bigg)$ with respect to $x$ at $x=1$. While the $n$-th order derivative vanishes exponentially as $\sim-\cos(n\pi/2)\text{e}^{-\pi l\lambda_T^2/4L_i^2}$ at a vary low temperature ($\lambda_T\gnsim L_i$), the same also vanishes exponentially as\footnote{The asymptotic form of the $n$th order derivative, for the case of high temperature, can be obtained from the use of the special form $2\sum_{j=1}^\infty\text{e}^{-a j^2}\cos(b j)+1=\vartheta_3(\frac{b}{2},\text{e}^{-a})=\sqrt{\frac{\pi}{a}}\text{e}^{-\frac{b^2}{4a}}+\sqrt{\frac{\pi}{a}}\sum_{k=1}^\infty\big(\text{e}^{-\frac{(2\pi k-b)^2}{4a}}+\text{e}^{-\frac{(2\pi k+b)^2}{4a}}\big)$ of the Poisson summation formula ($\sum_{j=-\infty}^\infty\delta(x-j)=\sum_{k=-\infty}^\infty\text{e}^{2\pi ikx}$) for $a>0$ and $b\in\text{reals}$.} $\sim\frac{1+(-1)^{n}}{2}[4L_i^2/\pi l\lambda_T^2]^{[2n+1]/2}\text{e}^{-\pi L_i^2/l\lambda_T^2}$ at a high temperature ($L_i\gg\lambda_T$). On the other hand, the first term $\frac{1}{2}\vartheta_3\big(0,\text{e}^{-\frac{l\beta\pi^2\hbar^2}{2mL_i^2}}\big)$ and the second term $-\frac{1}{2}\vartheta_3\big(\frac{\pi}{2},\text{e}^{-\frac{l\beta\pi^2\hbar^2}{2mL_i^2}}\big)$ of the square bracket in the above equation varies from $1/2$ to $\infty$ and $-1/2$ to $0$ respectively as $L_i$ varies from $0$ to $\infty$.  However, Eqn.(\ref{eqn:12b}) is an exact result for the total average number of particles. Chemical potential as well as the fugacity of the quantum gas can be exactly determined from this equation in terms of the temperature $T$ for fixed $\bar{N}$, $L_1$, $L_2$ and $L_3$. It is easy to check from the Bose-Einstein statistics that; as $T\rightarrow0$ is reached, the shifted chemical potential in units of $k_BT$ i.e. $[\mu-E_{1,1,1}]/k_BT$ reaches zero to the leading order $-1/\bar{N}$ for $\bar{N}\gg1$\footnote{For $T\rightarrow0$ almost all the bosons occupy the single-particle ground state resulting $\lim_{T\rightarrow0}\frac{1}{\text{e}^{[E_{1,1,1}-\mu]/k_BT}-1}=\bar{N}$. Here-from we get $[E_{1,1,1}-\mu]/k_BT=\ln(1+1/\bar{N})\simeq1/\bar{N}$.}. Incidentally, entropy per particle ($S/\bar{N}$) of the Bose system would not be zero, rather would be nonzero $k_B[\ln(1+\bar{N})]/\bar{N}$ due to the finite-size effect for $T\rightarrow0$ and $\bar{N}\gg1$ \cite{Pitaevskii}. However, deviation from the thermodynamic limit ($\bar{N}\rightarrow\infty$, $L_1\rightarrow\infty, L_2\rightarrow\infty, L_3\rightarrow\infty$ and $\bar{N}/[L_1L_2L_3]=constant$) gives rise to the finite-size effect which in-turn does not allow the Bose system to form the Bose-Einstein condensate unless $T\rightarrow0$ is reached. Reaching the condensation point ($T_c=0$) would be slower for the case of lower dimension, as because, the condensate does not form at a nonzero finite temperature in 2-D and 1-D even in the thermodynamic limit.    

\begin{figure}
\includegraphics[width=1 \linewidth]{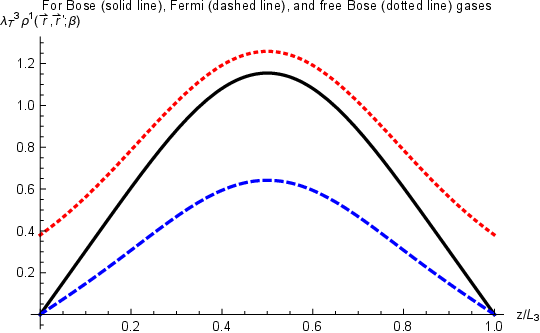}
\caption{Profile of the 1-particle density matrix-elements for the 3-D ideal Bose gas (solid line), Fermi gas (dashed line), which are free in $x-y$ plane and confined in the 1-D box of length $L_3$ along $z$-axis, for the fugacity $\bar{z}=0.8$, $\lambda_T/L_3=2/3$, $z'/L_3=1/2$, $x=x'$ and $y=y'$. Solid and dashed lines follow Eqn.(\ref{eqn:13}) for upper and lower signs respectively. The dotted line represents the same for 3-D free Bose gas \cite{Pitaevskii}
\label{fig2}}
\end{figure}

If boundaries along $x$ and $y$ axes are removed, i.e. if $L_1$ and $L_2$ are sent to $\infty$, then the system would be confined only along the $z$-axis. The one-particle density matrix in this situation would be
\begin{eqnarray}\label{eqn:13}
\rho^{1}(\vec{r},\vec{r}';\beta)=\sum_{l=1}^\infty(\pm1)^{l-1}\bar{z}^{l}\rho_f(\vec{r}_\perp,\vec{r}_\perp';l\beta)\rho(z,z';l\beta)
\end{eqnarray}
where $\vec{r}_\perp=x\hat{i}+y\hat{j}$, $\vec{r}_\perp'=x'\hat{i}+y'\hat{j}$,
\begin{eqnarray}\label{eqn:14}
\rho_f(\vec{r}_\perp,\vec{r}_\perp';\beta)=\frac{1}{\lambda_T^2}\text{e}^{-\frac{\pi |\vec{r}_\perp-\vec{r}_\perp'|^2 }{\lambda_T^2}}
\end{eqnarray}
is the density matrix for a free-particle in the $x-y$ plane \cite{Feynman}, and $\rho(z,z';\beta)$ is the density-matrix of the particle, similar to that in Eqn.(\ref{eqn:8}), for the bounded motion along the $z$-axis. The term with the first power of the fugacity $\bar{z}$ in the density matrix-element in Eqn.(\ref{eqn:13}) corresponds to the classical case. We plot the 1-particle density matrix-elements (Eqn.(\ref{eqn:13})) for 3-D ideal Bose gas and Fermi gas, and compare them with the one with the free-boundary for the Bose gas in FIG. \ref{fig2}. It is clear from the FIG. \ref{fig2} that the box-confinement reduces the spatial correlations in the system in comparison to that in the free quantum (Bose or Fermi) gas. The finite-size effect kills the long-ranged order in the 3-D ideal Bose gas, as expected, as the density matrix element has to vanish at the boundaries as clear in the FIG. \ref{fig2}. The dotted line would come arbitrary close to the solid line at $z=L_3/2$ for $\bar{z}\rightarrow0$. However, convergence of the series expansion of the 1-particle density matrix, as that in Eqn.(\ref{eqn:13}), would be slower for the case of the lower dimension so that contribution of the higher values of $l$ would not be negligible and it result stronger quantum fluctuations in the lower dimensions. This causes the density correlation to die out faster even between two arbitrary close points $\vec{r}$ and $\vec{r}'$ in the lower dimensions.
 
\begin{figure}
\includegraphics[width=1 \linewidth]{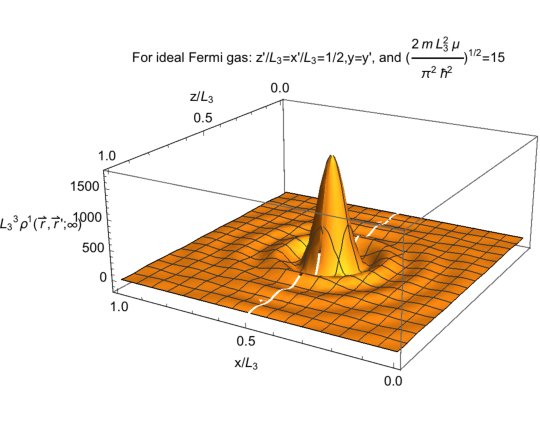}
\caption{Profile of the 1-particle density matrix-elements for the 3-D ideal Fermi gas for $T\rightarrow0$. The 3-D plot follows Eqn.(\ref{eqn:17}).
\label{fig3}}
\end{figure}

Alternating series expansion in Eqn.(\ref{eqn:13}) further reduces the spatial correlations in the Fermi gas in comparison to that in the Bose gas as clear in the FIG. \ref{fig2}. Eqn.(\ref{eqn:13}), for the Fermi gas, can be recast as
\begin{eqnarray}\label{eqn:16}
\rho^{1}(\vec{r},\vec{r}';\beta)&=&\sum_{j_3=1}^\infty\bigg[\int\frac{\text{e}^{i(\vec{r}_\perp'-\vec{r}_\perp)\cdot\vec{p}_\perp/\hbar}}{\text{e}^{\beta([p_\perp^2/2m+E_{j_3}]-\mu)}+1}\frac{\text{d}^2\vec{p}_\perp}{(2\pi\hbar)^2}\nonumber\\&&\times\frac{2}{L_3}\sin\bigg(\frac{j_3\pi z}{L_3}\bigg)\sin\bigg(\frac{j_3\pi z'}{L_3}\bigg)\bigg]\nonumber\\&=&\sum_{j_3=1}^\infty\bigg[\int_0^\infty\frac{2\pi J_0(|\vec{r}_\perp'-\vec{r}_\perp|p_\perp/\hbar)}{\text{e}^{\beta([p_\perp^2/2m+E_{j_3}]-\mu)}+1}\frac{p_\perp\text{d}p_\perp}{(2\pi\hbar)^2}\nonumber\\&&\times\frac{2}{L_3}\sin\bigg(\frac{j_3\pi z}{L_3}\bigg)\sin\bigg(\frac{j_3\pi z'}{L_3}\bigg)\bigg],
\end{eqnarray}
which can be further recast for $T\rightarrow0$, for which $\mu\ge\frac{\pi^2\hbar^2}{2mL_3^2}$ and $p_\perp$ ranges from $0$ to $p_{Fj_3}=\sqrt{2m\mu-\pi^2\hbar^2j_3^2/L_3^2}$, as~\footnote{Here $J_i$ represents Bessel function of order $i$ of the 1st kind.}
\begin{eqnarray}\label{eqn:17}
\rho^{1}(\vec{r},\vec{r}';\infty)&=&\frac{1}{\pi L_3}\sum_{j_3=1}^{\floor{\sqrt{\frac{2mL_3^2\mu}{\pi^2\hbar^2}}}}\bigg[\sin\bigg(\frac{j_3\pi z}{L_3}\bigg)\sin\bigg(\frac{j_3\pi z'}{L_3}\bigg)\nonumber\\&&\times \frac{p_{Fj_3}}{\hbar}\frac{J_1(|\vec{r}_\perp'-\vec{r}_\perp|p_{Fj_3}/\hbar)}{|\vec{r}_\perp'-\vec{r}_\perp|}\bigg].
\end{eqnarray}
We plot r.h.s. of Eqn.(\ref{eqn:17}) in FIG. \ref{fig3}. The oscillations in the density matrix-element in Eqn.(\ref{eqn:17}) are coming from the alternation of the sign of $\bar{z}^l$ in Eqn.(\ref{eqn:13}). Amplitude of the partial oscillations along $x$ and $y$ axes, however, is dying out hyperbolically as the system is not bounded along these two axes. The amplitude of the partial oscillations along the $z$-axis, however, are oscillating, as expected, as the system is bounded along this axis. The oscillatory amplitude oscillates rapidly except at around $z=z'$ if the Fermi energy (i.e. $\mu$ for $T\rightarrow0$) of the system increases; which further causes increase of the principal maximum of the density matrix-element of the system.

Spatial density correlation in the many-body system can be represented in terms of the 1-particle density matrix as $\nu(\vec{r},\vec{r}';\beta)=\pm\frac{|\rho^{1}(\vec{r},\vec{r}';\beta)|^2}{\rho^{1}(\vec{r},\vec{r};\beta)}$ \cite{Landau}. All the results we have got in this sections are thus useful to get the spatial density correlations in the many-body systems. These results can also be useful to get the quantum cluster expansions for the grand free energies of the many-body systems confined in the box geometries. 

\section{Cluster expansion for an ideal quantum gas in 3-D box}
Quantum cluster expansion of the grand free energy for a 3-D ideal quantum gas is given by \cite{Khan,Pathria}
\begin{eqnarray}\label{eqn:18}
\Omega=-k_BT\sum_{\nu=1}^\infty\frac{h_\nu\bar{z}^\nu}{\nu}
\end{eqnarray}
where 
\begin{eqnarray}\label{eqn:19}
h_\nu&=&(\pm1)^{\nu-1}\int..\int\rho(\vec{r}_1, \vec{r}_2; \beta)\rho(\vec{r}_2, \vec{r}_3; \beta)...\times\nonumber\\&&\rho(\vec{r}_{\nu}, \vec{r}_{1}; \beta)\text{d}^3\vec{r}_1...\text{d}^3\vec{r}_\nu~
\end{eqnarray}
is the cluster integral for $\nu$ indistinguishable particles in the system and $\vec{r}_i$ ($i=1,2,3...,\nu$) is the position vector for a particle in the cluster, and $\rho(\vec{r}_{i}, \vec{r}_{j}; \beta)$ is the single-particle density matrix element for the two positions $\vec{r}_{i}$ and $\vec{r}_{j}$ as defined in Eqn.(\ref{eqn:9}). The phase factor $(\pm1)^{\nu-1}$ in the above equation is appearing for the cyclic permutations of the bosons (upper sign) or fermions (lower) in the cluster. Eqn.(\ref{eqn:9}) leads to take the simplest form for the cluster integral $h_1$ as
\begin{eqnarray}\label{eqn:20}
h_1&=&\int_{0}^{L_1}\int_{0}^{L_2}\int_{0}^{L_3}\Pi_{i=1}^3\frac{1}{2L_i}\bigg[\vartheta_3\bigg(\frac{\pi[x_i-x_i]}{2L_i},\text{e}^{-\beta\frac{\pi^2\hbar^2}{2mL_i^2}}\bigg)\nonumber\\&&-\vartheta_3\bigg(\frac{\pi[x_i+x_i]}{2L_i},\text{e}^{-\beta\frac{\pi^2\hbar^2}{2mL_i^2}}\bigg)\bigg]\text{d}x_1\text{d}x_2\text{d}x_3\nonumber\\&=&\sum_{j_1,j_2,j_3=1}^{\infty,\infty,\infty}\text{e}^{-\beta\frac{\pi^2\hbar^2}{2m}\big(\frac{j_1^2}{L_1^2}+\frac{j_2^2}{L_2^2}+\frac{j_3^2}{L_3^2}\big)}.
\end{eqnarray}
Fourier series expansion of the Jacobi (elliptic) theta functions, as well as the Fourier series expansion shown in Eqn.(\ref{eqn:7}) for the density matrix-element for a particle in 1-D box, has been used to get the final expression of $h_1$ which by definition is the partition function for a single particle in the 3-D box. It can be shown from the orthonormality of the single-particle energy eigenstates that, two indistinguishable particles in the different energy eigenstates do not contribute to the cluster integral. Thus, by the definition, cluster integral for the two or more number of indistinguishable particles would be the partition function for the same number of particles always in the same single-particle energy eigenstate. Thus we get
\begin{eqnarray}\label{eqn:21}
h_\nu&=&\sum_{j_1,j_2,j_3=1}^{\infty,\infty,\infty}(\pm1)^{\nu-1}\text{e}^{-\nu\beta\frac{\pi^2\hbar^2}{2m}\big(\frac{j_1^2}{L_1^2}+\frac{j_2^2}{L_2^2}+\frac{j_3^2}{L_3^2}\big)}\nonumber\\&=&(\pm1)^{\nu-1}\Pi_{i=1}^3\frac{\vartheta_3\big(0,\text{e}^{-\frac{\nu\beta\pi^2\hbar^2}{2mL_i^2}}\big)-1}{2}
\end{eqnarray}
for $\nu=1,2,3,...$. The quantum cluster integral in the thermodynamic limit ($L_1,L_2,L_3\rightarrow\infty$), however, takes the form from to the leading order in $1/L_i$s from the above equation as $h_{\nu}^{(t-l)}=(\pm1)^{\nu-1}\frac{L_1L_2L_3}{\lambda_T^3\nu^{3/2}}$ \cite{Feynman}. We plot the quantum cluster integral in units of that in the thermodynamic limit in FIG. \ref{fig6} (inset) for $\nu=3$. However, now we get the quantum cluster expansion of the ideal quantum (Bose or Fermi) gas in the 3-D box, by recasting Eqn.(\ref{eqn:18}), as 
\begin{eqnarray}\label{eqn:22}
\Omega=-k_BT\sum_{\nu=1}^\infty(\pm1)^{\nu-1}\frac{\bar{z}^\nu}{\nu}\Pi_{i=1}^3\frac{\vartheta_3\big(0,\text{e}^{-\frac{\nu\beta\pi^2\hbar^2}{2mL_i^2}}\big)-1}{2}.~
\end{eqnarray}
Average number of particles ($\bar{N}=-\frac{\partial\Omega}{\partial\mu}|_{T,L_1L_2L_3}$), on the other hand can be calculated from Eqn.(\ref{eqn:22}) as
\begin{eqnarray}\label{eqn:23}
\bar{N}=\sum_{\nu=1}^\infty(\pm1)^{\nu-1}\bar{z}^\nu\Pi_{i=1}^3\frac{\vartheta_3\big(0,\text{e}^{-\frac{\nu\beta\pi^2\hbar^2}{2mL_i^2}}\big)-1}{2}.
\end{eqnarray}
Let us now define the generalized pressure given by the quantum gas on a wall of the system as $p=\frac{-\Omega}{L_1L_2L_3}$ which would be the true pressure in the thermodynamic limit. Thermodynamic behaviour of the ideal quantum gas  can now be extracted from Eqns. (\ref{eqn:22}) and (\ref{eqn:23}).

\begin{figure}
\includegraphics[width=1 \linewidth]{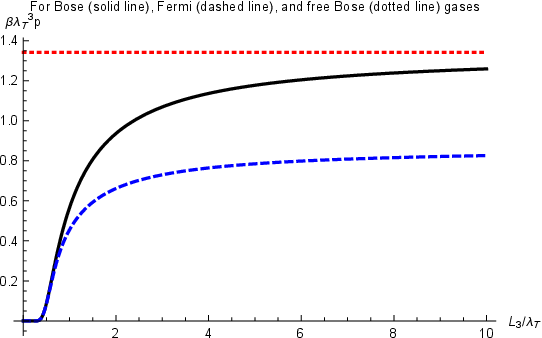}
\caption{Length dependence of the equation of states for the 3-D ideal Bose gas (solid line), Fermi gas (dashed line) and free Bose gas for $L_1\rightarrow\infty$, $L_2\rightarrow\infty$ and $\bar{z}=1$. The 1st two plots follow Eqn.(\ref{eqn:24}) for positive and negative signs respectively. 
\label{fig4}}
\end{figure}

Let us now remove the boundaries along the $x$ and $y$ axes so that $L_1\rightarrow\infty$, $L_2\rightarrow\infty$, and the quantum gas remains bounded only along the $z$ axis. In this situation the summations in Eqns. (\ref{eqn:22}) and (\ref{eqn:23}) over $j_1$ and $j_2$ can be replaced by integrations from $0$ to $\infty$ without any error by virtue of Poisson summation formula. Thus we get the equation of state of the system, i.e. the pressure given by the quantum gas to either of the walls situated at $z=0$ and $z=L_3$, as
 \begin{eqnarray}\label{eqn:24}
p=\frac{k_BT}{L_3\lambda_T^2}\sum_{\nu=1}^\infty(\pm1)^{\nu-1}\frac{\bar{z}^\nu}{\nu^2}\frac{\big[\vartheta_3\big(0,\text{e}^{-\nu\beta\frac{\pi^2\hbar^2}{2mL_3^2}}\big)-1\big]}{2}.~
\end{eqnarray}
This is quantum cluster expansion of the equation of state for ideal Bose and Fermi gases in the box geometry. As we mentioned before, the upper sign corresponds to the ideal Bose gas and the lower sign corresponds to the ideal Fermi gas. We plot the equation of state in Eqn. (\ref{eqn:24}) for ideal Bose and Fermi gases for fixed fugacity, and compare with the the case of the free Bose gas in FIG. \ref{fig4}. The solid line approaches the result $\lambda_T^3 p/k_BT=g_{5/2}(\bar{z})$, where $g_{5/2}$ is the Bose integral \cite{Pathria}, for the 3-D free Bose gas as $L_3/\lambda_T$ tends to $\infty$. The dashed line approaches the result $\lambda_T^3\beta p=f_{5/2}(\bar{z})$, where $f_{5/2}$ is the Fermi integral \cite{Pathria}, for the 3-D free Fermi gas as $L_3/\lambda_T$ tends to $\infty$.  It is clear from the FIG. \ref{fig4} that, finiteness of system causes less pressure given to the wall with respect to that given by a free gas though finiteness causes more energy to the system which is not probabilistically favoured in thermal equilibrium. This pressure is exponentially small for $\lambda_T\gnsim L_3$ as clear from the FIG. \ref{fig4} too. The finite system effectively behaves like a 2-D system in such a low temperature regime. This pressure, however, is further reduced in Fermi gas for odd permutation effect in clusters for even $\nu$s. Casimir-like effect can also be studied from Eqn. (\ref{eqn:24}) for the finiteness of the system only along the $z$ axis \cite{Biswas2007} apart from studying typical finite-size effect on the many-body system. 

Onset of Bose-Einstein condensation below the condensation point $T_c$, on the other hand, needs all the cluster integral with equal importance. This is not possible in a finite system as $\mu-E_{1,1,1}$ does not vanish below the bulk condensation point ($T_0$); rather the cluster integrals vanish for $T\rightarrow0$ for quantum gases in the box geometry as clear from the FIG. \ref{fig6} (inset).

\section{Cluster expansion for a non-ideal quantum gas in a closed rectangular cylinder}
We already have understood that finite-size effect a Bose system to form a condensate at a non-zero finite temperate.  The Bose condensate irrespective of its dimension, however, would be robust only for $T\rightarrow0$ though the finite-size effect is even stronger in lower dimensions. Let the Bose gas spin $0$ particles be interacting with short-ranged pair-potential energy $V(|\vec{r}_2-\vec{r}_1)|)$ in a 3-D closed rectangular cylindrical box, so that the many-body Hamiltonian of the system can be written, in usual notation, as \cite{Dalfovo}
\begin{eqnarray}\label{eqn:25}
\hat{H}&=&\int\text{d}^3\vec{r}\hat{\psi}^{\dagger}(\vec{r})\bigg(-\frac{\hbar^2}{2m}\nabla^2\bigg)\hat{\psi}(\vec{r})+\nonumber\\&&\frac{1}{2}\int\int \text{d}^3\vec{r}'\text{d}^3\vec{r}\hat{\psi^{\dagger}}(\vec{r})\hat{\psi^{\dagger}}(\vec{r}')V(|\vec{r}-\vec{r}'|)\hat{\psi}(\vec{r}')\hat{\psi}(\vec{r})~~~~~
\end{eqnarray}
where $V(|\vec{r}-\vec{r}')|)=g_3\delta^3(|\vec{r}-\vec{r}'|)$\footnote{This is the limiting case of Fermi-Huang potential under Born approximation for low energy scattering \cite{Bhattacharya}.} is considered to be a short-ranged pair-potential energy, $g_3=\frac{4\pi\hbar^2a_s}{m}$ is the coupling constant, and $a_s$ is the s-wave scattering length which is positive for repulsive interactions. The system would not be stable beyond a critical number of particles for $g_3<0$ which we are not considering in our analysis \cite{Biswas2009}. Let us further consider that, $L_1\ll L_3$ and $L_2\ll L_3$ so that low-lying excitations only along $z$-direction are probabilistically favoured for low temperatures (for which $\lambda_T\gg L_1$ and $\lambda_T\gg L_2$). The system behaves like a quasi 1-D system ($0\le z\le L_3$) in this situation. If the average number of indistinguishable bosons in the grandcanonical ensemble (for temperature $T$ and chemical potential $\mu$) be $\bar{N}$, then the 1-particle (low-lying) excitations ($\psi_j(z);~~j=1,2,3,...$) follow from the time-independent non-linear Schrodinger equation \cite{Carr}
\begin{eqnarray}\label{eqn:26}
\bigg[-\frac{\hbar^2}{2m}\frac{\text{d}^2}{\text{d} z^2}+g_1\bar{N}|\psi_j(z)|^2\bigg] \psi_j(z)=\bar{E}_j\psi_j(z)
\end{eqnarray}
with the 1-particle mean field energy eigenvalue \cite{Carr}
\begin{eqnarray}\label{eqn:27}
\bar{E}_j=\frac{2\hbar^2j^2}{mL_3^2}(1+q_j)K^2(q_j),
\end{eqnarray}
where $K(q_j)=\frac{\pi}{2}\big[1+q_j/4+9q_j^2/64+...\big]$ is complete elliptic integral of the first kind of the argument $q_j=\frac{g_1\bar{N}|\psi_j(L_3/2j)|^2}{2E_j}/\big[1-\frac{g_1\bar{N}|\psi_j(L_3/2j)|^2}{2E_j}\big]$ (for $0<q_j<1/2$ \cite{SB}), $g_1=\frac{2\hbar^2}{ma_s}$ is the coupling constant for elastic scattering in 1-D, and \cite{Carr}
\begin{eqnarray}\label{eqn:28}
\psi_j(z)=\sqrt{\frac{\hbar^2q_j}{mL_3^2g_1\bar{N}}}\big[2jK(q_j)\big]\text{sn}\bigg(\frac{2jK(q_j)z}{L_3},q_j\bigg)
\end{eqnarray}
is the solution (Jacobi elliptic function) to Eqn.(\ref{eqn:26}) for the Dirichlet boundary conditions. One-particle density matrix, for the interacting Bose system, thus takes the form, from the definition, as
\begin{eqnarray}\label{eqn:29}
\rho^1(z,z';\beta)&=&\sum_{j=1}^{\infty}\bigg[\frac{\hbar^2q_j4j^2K^2(q_j)}{mL_3^2g_1\bar{N}}\frac{1}{\text{e}^{\beta[\bar{E_j}-\mu]}-1}\nonumber\times\\&&\text{sn}\bigg(\frac{2jK(q_j)z}{L_3},q_j\bigg)\text{sn}\bigg(\frac{2jK(q_j)z'}{L_3},q_j\bigg)\bigg].~~~~~
\end{eqnarray}
Quantum cluster integral for the cycle of length $\nu=1,2,3,...$ can now be easily written, for the interacting Bose system, by looking at the Eqn.(\ref{eqn:21}), as
\begin{eqnarray}\label{eqn:30}
h_\nu=\sum_{j=1}^\infty\text{e}^{-\nu\beta\bar{E}_j}.
\end{eqnarray}
It is easy to check, that, all these results are matching with the cases of ideal Bose gas in the 1-D box for $g_1\rightarrow0$.

Let us now consider the case of interacting Fermi gas of spin 1/2 particles (of 1:1 components of $s_z=1/2$ and $-1/2$) in the same situation as above. Pair interactions are now possible between the particles of $s_z=1/2$ and $s_z=-1/2$ as effect of other direct pair interactions are cancelled due to that of exchange pair interactions \cite{Biswas2012}. One-particle density matrix and the cluster integral can now be defined only for one species of the spin component. Thus $\bar{N}$ in Eqns.(\ref{eqn:26}) - (\ref{eqn:30}) would be replaced by $\bar{N}/2$ for the interacting Fermi system. In addition the Bose-Einstein statistics ($\bar{n}_j=\frac{1}{\text{e}^{\beta[\bar{E_j}-\mu]}-1}$) in Eqn.(\ref{eqn:29}) is to be replaced by the Fermi-Dirac statistics ($\bar{n}_j=\frac{1}{\text{e}^{\beta[\bar{E_j}-\mu]}+1}$) and a phase factor $(-1)^{\nu-1}$ would be multiplied in the r.h.s. of Eqn.(\ref{eqn:30}).

Density matrix elements have been obtained for systems in box geometries revealing the breaking of the translational symmetry in the finite geometry. Quantum cluster expansions have also been obtained revealing effects of both the 1st quantization and the 2nd quantization on the cluster integrals. The effect of the 1st quantization can be killed in the thermodynamic limit which sends size of the system to infinity. The same, however, can also be studied for the harmonically trapped Bose and Fermi gases.

\section{Quantum cluster expansion for 3-D Bose or Fermi gas in the harmonically trapped geometry}
While the quantum cluster expansion for the 3-D Bose or Fermi gas has not been evaluated, except the approximated one evaluated for strongly interacting Fermi gas in the classical (or high temperature) regime \cite{Hu}, exact result for the statistical mechanical density matrix element was obtained as \cite{Feynman}  
\begin{eqnarray}\label{eqn:31}
\rho(x,x';\beta)&=&\sqrt{\frac{m\omega_x}{2\pi\hbar\sinh(\beta\hbar\omega_x)}}\nonumber\\&&\times\text{e}^{-\frac{m\omega_x}{2\hbar\sinh(\beta\hbar\omega_x)}[(x^2+x'^2)\cosh(\beta\hbar\omega_x)-2xx']}~~~~~
\end{eqnarray}
where $\omega_x$ is the angular frequency of oscillation along the $x$-axis with the center at $x=0$ for the harmonic oscillator. This equation for the harmonic oscillator counters the Eqn.(\ref{eqn:8}) obtained for the particle in the 1-D box. Then density matrix element for a 3-D harmonic oscillator, with the angular frequencies $\omega_x$, $\omega_y$ and $\omega_z$ for oscillations along $x$, $y$ and $z$ axes respectively, can be written in the separable form as $\rho(\vec{r},\vec{r}';\beta)=\rho(x,x';\beta)\rho(y,y';\beta)\rho(z,z';\beta)$ where each of $\rho(x,x';\beta)$, $\rho(y,y';\beta)$ and $\rho(z,z';\beta)$ has same form as in Eqn.(\ref{eqn:31}) along with the corresponding variables. 

\subsection{One-particle density matrix for ideal Bose or Fermi gas in harmonic traps}
One-particle density matrix element for ideal Bose or Fermi gas in such a harmonic trap can be written in the same form Eqn.(\ref{eqn:12}) as in with this single-particle density matrix $\rho(\vec{r},\vec{r}';\beta)$. Thus the one-particle density matrix element takes the form for the trapped ideal quantum gas as
\begin{eqnarray}\label{eqn:32}
\rho^{1}(\vec{r},\vec{r}';\beta)=\sum_{l=1}^\infty(\pm1)^{l-1}\bar{z}^{l}\rho(x,x';l\beta)\rho(y,y';l\beta)\rho(z,z';l\beta)\nonumber\\
\end{eqnarray}
where the fugacity $\bar{z}$ of the trapped quantum gas. The evaluation of the 1-particle density matrix in Eqn.(\ref{eqn:32}) would be of simple form for isotropic trap ($\omega_x=\omega_y=\omega_z=\bar{\omega}$) as
\begin{eqnarray}\label{eqn:33}
\rho^{1}(\vec{r},\vec{r}';\beta)&=&\sum_{l=1}^\infty\bigg[\frac{m\bar{\omega}}{2\pi\hbar\sinh(l\beta\hbar\bar{\omega})}\bigg]^{3/2}(\pm1)^{l-1}\bar{z}^{l}\nonumber\\&&\times\text{e}^{-\frac{m\bar{\omega}}{2\hbar\sinh(l\beta\hbar\bar{\omega})}[(r^2+r'^2)\cosh(l\beta\hbar\bar{\omega})-2\vec{r}\cdot\vec{r}']}~~~~~
\end{eqnarray}
It is easy to check from the above equation that, although the rotation symmetry of the 1-particle density matrix of the ideal quantum gas is preserved in the harmonically trapped geometry, yet the translation symmetry of the same is broken as a finite-size effect. The translation symmetry can be restored only for $\bar{\omega}\rightarrow0$.  
Total average number of particles can be obtained from Eqn.(\ref{eqn:33}) as $\bar{N}=\int\rho^{1}(\vec{r},\vec{r};\beta)\text{d}^3\vec{r}=\sum_{l=1}^\infty[1/2\sinh(l\beta\hbar\bar{\omega}/2)]^3(\pm)^l\bar{z}^l$. This result for anisotropic trap would be
\begin{eqnarray}\label{eqn:34}
\bar{N}=\sum_{l=1}^\infty\frac{1}{2\sinh(\frac{l\beta\hbar\omega_x}{2})}\frac{1}{2\sinh(\frac{l\beta\hbar\omega_y}{2})}\frac{1}{2\sinh(\frac{l\beta\hbar\omega_z}{2})}(\pm)^l\bar{z}^l.\nonumber\\
\end{eqnarray}
The fugacity as well as the chemical potential are to be determined from Eqn.(\ref{eqn:34}). 

\begin{figure}
\includegraphics[width=1 \linewidth]{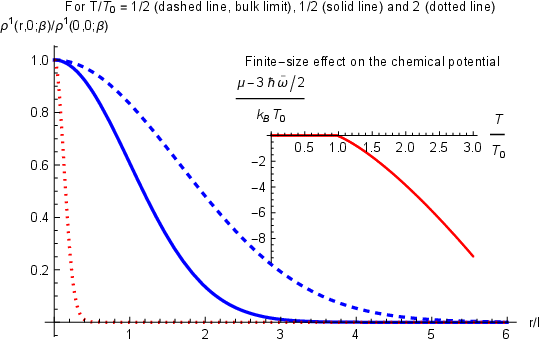}
\caption{Plots for finite-size effect on the 1-particle density matrix element for the harmonically trapped ideal Bose gas of $^{87}$Rb atoms in a trap of finite-size. Plots follow Eqn.(\ref{eqn:33}) for $\bar{N}=40\times10^3$ and $\omega_x=\omega_y=\omega_z=\bar{\omega}=1140~$Hz. The inset follows from Eqn.(\ref{eqn:34}) for the same parameters. Here $l=\sqrt{\hbar/M\bar{\omega}}\simeq0.796744~\mu$m represents the finite length-scale of the trap, and $T_0=\frac{\hbar\bar{\omega}}{k_B}(\bar{N}/\zeta(3))^{1/3}\simeq280~$nK is the Bose-Einstein condensation point in the thermodynamic limit \cite{Ensher}. The dashed line is obtained in the thermodynamic limit for $T/T_0=1/2$, and the solid and dotted lines are obtained for the finite-size of the trap for $T/T_0=1/2$ and $2$ respectively. 
\label{fig5}}
\end{figure}

We show the finite-size effect on the chemical potential in the inset of the FIG. \ref{fig5}. However, the Bose-Einstein condensate does not form for the trapped Bose gas (upper sign) below the condensate point $T_c$ unless the thermodynamic limit is reached ($N\rightarrow\infty$, $\omega_x\rightarrow0$, $\omega_y\rightarrow0$, $\omega_z\rightarrow0$, and $N\bar{\omega}^3=constant$). The fugacity can be expressed in terms of the temperature, geometric mean of the angular frequencies ($\bar{\omega}=\sqrt[3]{\omega_x\omega_y\omega_z}$) and the total average number of particles $\bar{N}$ in the thermodynamic limit as described in Ref.\cite{Biswas2012b}. Deviation from the thermodynamic limit gives rise to the finite-size effect which in-turn does not allow the system to form the Bose-Einstein condensate unless $T\rightarrow0$ is reached. The result (i.e. the dotted line in the FIG. \ref{fig5}) in the thermodynamic limit was obtained for $T/T_0=1/2$ from the same Eqn.(\ref{eqn:33}) with the further consideration of the leading orders of $\bar{\omega}$ and proportionate contributions from the Bose-Einstein condensate and the thermal cloud. We plot Eqn.(\ref{eqn:33}) in FIG. \ref{fig5} for isotropic trap and with the fugacity determined from Eqn.(\ref{eqn:34}) within the graphical method. It is clear from the FIG. \ref{fig5} that, the finite-size effect reduces the spatial density correlation in the system, and the spatial density correlation is less at a higher temperature as expected. It is interesting to note that a small change in chemical potential from its bulk value below the bulk condensation point ($T_0$) is sensitive to the macroscopic property of the system, and causes a large amount of the difference in the 1-particle density matrix element as clear from the  FIG. \ref{fig5}. Onset of off-diagonal long reneged order is not possible even in the thermodynamic limit of the trapped system. Observation of the same needs a 3-D free Bose gas \cite{Pitaevskii}.

\subsection{Quantum cluster expansion for ideal Bose or Fermi gas in harmonic traps}
Quantum cluster integral for the cycle of the length $\nu$, on the other hand, can be written for the harmonically trapped ideal quantum gas, in the spirit of Eqn.(\ref{eqn:21}), as  
\begin{eqnarray}\label{eqn:35}
h_\nu&=&\sum_{j_1,j_2,j_3=0}^{\infty,\infty,\infty}(\pm1)^{\nu-1}\text{e}^{-\nu\beta[(j_1+1/2)\hbar\omega_x]}\nonumber\\&&\times\text{e}^{-\nu\beta[(j_2+1/2)\hbar\omega_y+(j_3+1/2)\hbar\omega_z]}\nonumber\\&=&(\pm1)^{\nu-1}\Pi_{i=1}^3\frac{1}{2\sinh(\nu\beta\hbar\omega_i/2)}
\end{eqnarray}
where $\omega_x=\omega_1$, $\omega_y=\omega_2$ and $\omega_z=\omega_3$. The quantum cluster integral in the thermodynamic limit ($\omega_1,\omega_2, \omega_3\rightarrow0$), however, takes the form to the leading order in the angular frequencies from the above equation as $h_{\nu}^{(t-l)}=(\pm1)^{\nu-1}\frac{1}{(\beta\hbar\bar{\omega})^3\nu^{3}}$ \cite{Pitaevskii}. 

\begin{figure}
\includegraphics[width=1 \linewidth]{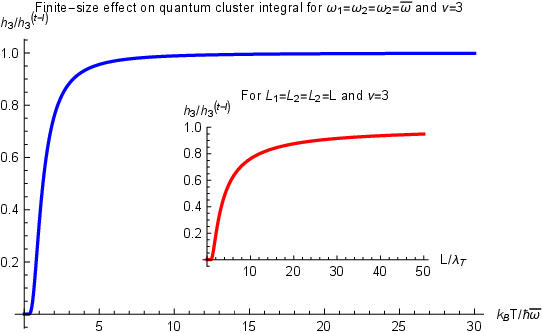}
\caption{Finite-size effect on the quantum cluster integral. Plot for the quantum (Bose or Fermi) gas follows Eqn.(\ref{eqn:35}) for the parameters as mentioned in the plot-label. The plot in the inset represents the same for the box geometry and follows Eqn.(\ref{eqn:21}) for the parameters as mentioned in the plot-label. 
\label{fig6}}
\end{figure}

Quantum cluster expansion of the grand free energy for the 3-D ideal quantum gas in the harmonic trap can be written by following its definition (Eqn.(\ref{eqn:18})) for the above cluster integrals as
\begin{eqnarray}\label{eqn:36}
\Omega=-k_BT\sum_{\nu=1}^\infty(\pm1)^{\nu-1}\frac{\bar{z}^\nu}{\nu}\Pi_{i=1}^3\frac{1}{2\sinh(\nu\beta\hbar\omega_i/2)}.
\end{eqnarray}
All the relevant thermodynamic variables can be obtained from this grand free energy for the harmonically trapped Bose or Fermi gas. For non-ideal case, on the other hand, the cluster integrals take the similar form as in Eqn.(\ref{eqn:35}), as described in Eqn.(\ref{eqn:30}) for the 1-D box, with the $(j_1+1/2)\hbar\omega_x+(j_2+1/2)\hbar\omega_y+(j_3+1/2)\hbar\omega_z$ be replaced by the energy eigenvalues of the Hartree-Fock equation \cite{Pitaevskii}. However, we plot the quantum  integral in units of that in the thermodynamic limit in FIG. \ref{fig6} for $\nu=3$. While the quantum cluster integrals for Bose and Fermi gases are having $\pi$-phase difference for even values of the cycle-length, they are same for odd values of the cycle-length ($\nu$). It is clear from the FIG. \ref{fig6} that quantum cluster integrals reach their respective thermodynamic limiting values either at the very large temperature or at the very large system size.

\section{Conclusion}
To conclude -- we have analytically obtained 1-particle density matrices, from the scratch points, for the ideal Bose and Fermi gases in both the 3-D box geometries and the harmonically trapped geometries for the entire range of temperature. We have obtained quantum cluster expansions of the grand free energies in closed forms for these systems. We have shown how the density matrix formalism can be adopted for obtaining the cluster integral for a system of finite-size, such as, ideal Bose or Fermi gas in a 3-D box or in a harmonic trap. Thermodynamics of the  Bose and Fermi gases in the restricted geometries can be studied from the quantum cluster expansions obtained by us. We also have obtained analytic results for non-deal cases, in particular, for interacting Bose and Fermi gases confined in closed rectangular cylinders. 

Eqn.(\ref{eqn:21}) and (\ref{eqn:35}) are our key results for the quantum cluster integral ($h_\nu$) for the cycle of length $\nu$ for an ideal quantum (Bose or Fermi) gas in a 3-D box or harmonic trap respectively. These two equation though look different, they have a generic form $h_\nu=(\pm1)^{\nu-1}\sum_{j_1,j_2,j_3}\text{e}^{-\nu\beta E_{j_1,j_2,j_3}}$ where the summation spreads over all the single-particle energy eigenstates with the eigenvalues ($\{E_{j_1,j_2,j_3}\}$). The generic form appears to be the canonical partition function for a composite particle of indistinguishable bosons  or fermions in a cluster of size $\nu$. Surprisingly, neither the special form of the quantum cluster integrals nor their generic form were obtained before us for a quantum gas in the box or harmonically trapped geometry. Proof of the generic form is given in the Appendix-\ref{A}. We are now theoreming that, the generic form of the quantum cluster integral ($h_\nu$) for a cluster of size $\nu$ of any system of indistinguishable bosons (upper sign) or fermions (lower sign) in thermodynamic equilibrium would be $(\pm)^{\nu-1}$ times the canonical partition function ($Z(\nu\beta)$) for a composite particle composed of the indistinguishable bosons or fermions in a cluster of size $\nu$ at a temperature $T=1/k_B\beta$. The phase factor $(\pm1)^{\nu-1}$ is appearing for the cyclic permutations of the bosons (upper sign) or fermions (lower) in the cluster. For indistinguishable particles of an ideal gas in a box, $Z(\nu\beta)$ can be interpreted as the canonical partition function for the composite particle of reduced mass $m/\nu$ of the $\nu$ constituent particles of mass $m$ each in the cluster at a temperature $T=1/k_B\beta$.

Our theoretical results are exact, and are directly useful for understanding finite-size effects on quantum cluster expansion of Bose and Fermi gases in the restricted geometries. Our results would be relevant in the context of experimental study of spatial correlations in ultra-cold systems of dilute Bose and Fermi gases of alkali atoms (i) in 3-D magneto-optical box traps with quasi-uniform potential around the center \cite{Gaunt}, and (ii) in 3-D harmonic traps \cite{Ensher,Jin}. Our method of obtaining the quantum cluster expansion with the realization of the quantum cluster integral as the partition function of a composite particle and having a phase, would be a rigorous approach for studying finite-size effects in statistical mechanics.

Here, by density matrix, we mean: statistical mechanical unnormalized density matrix for systems in thermodynamic equilibrium. Replacing $\beta$ by $it/\hbar$, where $t$ is the time taken by a particle in the system to reach $\vec{r}'$ starting from the position $\vec{r}$ at $t=0$, we get the quantum mechanical density matrix-elements for all the cases.  

Our results can be generalized for rotating traps without any difficulties as energy eigenvalues for the harmonic oscillator do not change much if the trap rotates slowly about the symmetric axis \cite{Fetter}. The case of the rotation would be more interesting for the fast rotation, specially, when angular speed of rotation becomes same as that of the (angular) trap frequency resulting infinite degeneracy of the ground state. All our results can be generalized for non-ideal cases of 3-D Bose and Fermi systems in the box geometries or in the harmonically trapped geometries within the Hartree-Fock approximations. We are keeping these tasks as open problems to the scientific community. 

\section*{Acknowledgement}
S. Dey, S. Basu and D. Banerjee acknowledge partial financial support of the UGC-Networking Resource Centre, School of Physics, University of Hyderabad, India (July-September, 2019). P. Manchala acknowledges financial support (JRF) of the CSIR, Govt. of India. S. Biswas acknowledges partial financial support of the SERB, DST, Govt. of India under the EMEQ Scheme [No. EEQ/2019/000017]. We thank the anonymous reviewers for their thorough review, and highly appreciate their comments and suggestions which significantly contributed to improving the quality of the presentation.

\appendix
\section{Proof of the generic form of the quantum cluster integral}
\label{A}

The cluster expansion of the grand free energy for ideal quantum gas is given by the expression
\begin{equation}\label{A1}
\Omega = - k_B T \sum_{\nu = 1}^{\infty} \frac{h_\nu \bar{z}^\nu}{\nu}
\end{equation}
where $h_\nu$ is defined as the quantum cluster integral of cycle-length $\nu$ as
\begin{eqnarray}\label{A2}
h_\nu &=& (\pm 1)^{\nu-1}\big[\prod_{j=1}^{\nu} \int\text{d}^3\vec{r}_i \big]\rho(\vec{r}_1, \vec{r}_2,\beta)\rho(\vec{r}_2, \vec{r}_3,\beta)...\nonumber\\&&\rho (\vec{r}_{j-1}, \vec{r}_{j},\beta)...\rho  (\vec{r}_\nu,r_1,\beta)
\end{eqnarray}
like that in Eqn.(\ref{eqn:19}). To simplify this expression we write the integral over density matrices in the operator notation. We then note that, the identity can be decomposed as a sum of projection operators over a complete set of orthonormal states, e.g. $\mathbb{1}=\sum_{j=0}^\infty|\psi_j\rangle\langle\psi_j|$ for the 1-D harmonic oscillator. We go on with this procedure in position representation i.e. $\mathbb{1}=\int|\vec{r}_j\rangle\langle\vec{r}_j|\text{d}^3\vec{r}_j$ for all $r_j$s with the change of notation $\rho(\vec{r}_{j}, \vec{r}_{j+1},\beta)\rightarrow\rho(\vec{r}_{j}, \vec{r}_{j+1})=\braket{\vec{r}_{j}|\hat{\rho}|\vec{r}_{j+1}}$ for $j=1,2,...,\nu$ \& $\vec{r}_{\nu+1}=\vec{r}_{1}$, as follows.
\begin{eqnarray}\label{A3}
\frac{h_\nu}{(\pm 1)^{\nu-1}}&=&\int...\int\rho (\vec{r}_1, \vec{r}_2)...\rho(\vec{r}_{\nu-1}, \vec{r}_{\nu})\rho(\vec{r}_\nu, \vec{r}_1)\text{d}^3\vec{r}_1...\text{d}^3\vec{r}_\nu\nonumber\\&=&\big[ \prod_{j=1}^{\nu}\int\text{d}^3\vec{r}_j \big]\braket{\vec{r}_{1}|\hat{\rho}|\vec{r}_{2}}...\braket{\vec{r}_{j-2}|\hat{\rho}|\vec{r}_{j-1}}\braket{\vec{r}_{j-1}|\hat{\rho}|\vec{r}_{j}}\nonumber\\&&...\braket{\vec{r}_{\nu}|\hat{\rho}|\vec{r}_{1}}\nonumber\\&=&\int\text{d}\vec{r}_1 \braket{\vec{r}_1|\hat{\rho}^\nu|r_1} = \text{Tr}.(\hat{\rho}^\nu)=\text{Tr}.(\text{e}^{-\nu\beta\hat{H}})\nonumber\\&=&\sum_{j_1, j_2, j_3}\text{e}^{-\beta \nu E_{j_1, j_2, j_3}},
\end{eqnarray}
where $E_{j_1, j_2, j_3}$ is the energy eigenvalue in the single-particle sate  $\psi_{j_1, j_2, j_3}(\vec{r})$ and the summation over $j_1, j_2, j_3$ are taken over the complete and orthonormal set of the single-particle states. Hence, we obtain the quantum cluster integral for the cycle of the length $\nu$ as 
\begin{equation}\label{A4}
h_\nu = (\pm 1)^{\nu-1}\sum_{j_1, j_2, j_3}\text{e}^{-\beta \nu E_{j_1, j_2, j_3}}=(\pm 1)^{\nu-1}Z(\nu\beta)
\end{equation}
where $Z(\nu\beta)$ is the canonical partition function for a composite particle composed of the $\nu$ indistinguishable constituent particles in the cluster of size $\nu$ at a temperature $T=1/k_B\beta$. It is to be noted that, the energy is an extensive variable at least for the ideal gas of particles. So, the energy of the composite particle in a given energy eigenstate $\ket{\psi_{j_1, j_2, j_3}}$ is $\nu$ times the energy of a constituent particle in the same eigenstate in the cluster of size $\nu$.
\end{document}